\documentclass[numreferences]{kluwer}
\newcommand{\be}{\begin{equation}}\newcommand{\ee}{\end{equation}}
\newcommand{\bea}{\begin{eqnarray}}\newcommand{\eea}{\end{eqnarray}}
\newcommand{\brr}{\begin{array}}\newcommand{\err}{\end{array}}
\newcommand{\ben}{\begin{enumerate}}\newcommand{\een}{\end{enumerate}}
\newcommand{\bib}{\bibitem}
\newcommand{\ba}{\begin{array}}
\newcommand{\ea}{\end{array}}

\def\lab{\label}\def\lan{\langle}
\def\lf{\left}\def\lrar{\leftrightarrow}

\def\non{\nonumber}\def\pa{\partial}\def\ran{\rangle}
\def\rar{\rightarrow}
\def\ri{\right}\def\ti{\tilde}
\def\al{\alpha}\def\bt{\beta}\def\ga{\gamma}\def\Ga{\Gamma}
\def\de{\delta}\def\De{\Delta}\def\ep{\epsilon}
\def\te{\theta}\def\Te{\Theta}
\def\si{\sigma}
\def\om{\omega}
\def\AB{{_{A,B}}}\newcommand{\mlab}[1]{\label{#1}}
\def\T{{_{T}}}
\def\AB{{_{A,B}}}\def\mass{{_{1,2}}}
\def\1{{_{1}}}\def\2{{_{2}}}
\def\bk{{\bf {k}}}\def\bx{{\bf {x}}}

\newcommand{\ide}{1\hspace{-1mm}{\rm I}}

\begin{document}
\begin{article}
\begin{opening}
\title{Quantum Field Theory of particle mixing and oscillations}
\author{Massimo  \surname{Blasone}}
\runningauthor{M.Blasone and G.Vitiello}
\runningtitle{QFT of particle mixing and oscillations}
\institute{The Blackett Laboratory, Imperial College London, London
SW7 2AZ, U.K. and \\
Institute f\"ur Theoretische Physik, Freie Universit\"at Berlin,
Arnimallee 14, D-14195 Berlin, Germany}

\author{Giuseppe Vitiello}
\institute{Dipartimento di
Fisica and INFN, Universit\`a di Salerno, 84100 Salerno, Italy}

\begin{abstract}
We report on recent results on the Quantum Field Theory of
mixed particles. The quantization procedure is discussed in detail, both for
fermions and for bosons and the unitary inequivalence of the flavor and
mass representations is proved.
Oscillation formulas exhibiting
corrections with respect to the usual quantum
mechanical ones are then derived.
\end{abstract}

\end{opening}

\section{Introduction}

The chapter of particle mixing and
 oscillations \cite{Cheng-Li}  is one of
the most important and fascinating in the book of
modern Particle
Physics. This is especially true after the recent experimental
results \cite{experiments} which finally confirm, after a long
search, the reality of neutrino oscillations \cite{Pont,Sol}:
this represents indeed the first clear evidence for physics beyond the
Standard Model.

Many unanswered questions about the
physics of particle mixing are however still there,
in particular from a theoretical
point of view. Apart from the problem of the origin of mixing and
of the small neutrino masses, difficulties arise already in the attempt
to find a proper mathematical setting for the description of
mixed particles in Quantum Field Theory (QFT).

It is indeed well known \cite{Barg} that mixing of states with
different masses is not allowed\footnote{See however
also Ref.\cite{Greenb}.} in non-relativistic Quantum Mechanics (QM).
In spite of this fact, the quantum mechanical treatment is the one usually
adopted for its simplicity and elegance. A review of the problems connected
with the QM of mixing and oscillations can be found in
Ref.\cite{Zralek}. Difficulties in the construction of the Hilbert
space for mixed neutrinos were pointed out in Ref.\cite{kimbook}.

Only recently \cite{BV95}-\cite{observables}
a consistent treatment of mixing and
oscillations in
QFT has been achieved and we report here on these results.

The main point of  our analysis \cite{BV95} consists in the
observation that a problem of representation (i.e. choice of the
Hilbert space) may arise when we start to mix fields with different
masses. This  has to do with the peculiar mathematical structure
of QFT, where unitarily inequivalent representations  of the
algebra of fields do exist \cite{Itz,Um1}: a classical example is
the one of theories with spontaneous breakdown  of symmetry. This
situation is in contrast to the one of QM, which deals with
systems with a finite number of degrees of freedom and where only
one Hilbert space is admitted (von Neumann theorem).

On this basis,
a careful analysis of the usual mixing transformations in QFT reveals
a rich non-perturbative  structure associated to the vacuum for mixed
particles, which
appears to be a condensate of particle-antiparticle pairs, both
for fermions and bosons.
The vacuum for the mixed fields
is a generalized coherent state \`a la Perelomov \cite{Per}.

The structure of flavor vacuum reflects into observable quantities:
exact oscillation formulas \cite{BHV99,bosonmix}
are derived in QFT exhibiting corrections
with respect to the usual QM ones.
We also show that a geometric phase
is associated to flavor  oscillations
\cite{berry}.

The material here presented  is organized in the following way:

In Section \ref{mixing}, the mixing transformations
are studied in QFT, both
for  fermions and bosons, in the case of two flavors.
The currents and charges for mixed fields
are also introduced and then used in Section \ref{oscill} to derive exact oscillation
formulas for charged fields (bosons and fermions). The
case of neutral fields is treated in Section \ref{neutral}.

The geometric phase for oscillating particles is studied in Section
\ref{berry}. In Section \ref{3flav} the case of three flavor mixing is
considered and the deformation of the associated algebra due to CP violation
is discussed. Finally, in Section \ref{spacemix}, a
space dependent oscillation formula
for neutrinos is derived using the relativistic flavor current.

\section{Mixing transformations in Quantum Field Theory}
\label{mixing}

In this Section we study the quantization of mixed fields
both for Dirac fermions   and  for charged bosons
\cite{BV95,lathuile,bosonmix}.
For simplicity, we limit ourselves to the case of two generations
(flavors) although the main results presented below have general validity
\cite{hannabuss}. Three flavor fermion mixing \cite{3flavors}
is discussed in \S\ref{3flav}.

\subsection{Fermion mixing}
\label{fermionmix}

Let us consider\footnote{We refer to neutrinos, but the discussion is
clearly valid for any Dirac fields.} two flavor fields  $\nu_e$,
$\nu_{\mu}$. The mixing relations are \cite{Pont}
\bea\non
\nu_{e}(x) &=&  \cos\te \;\nu_{1}(x)  \, +\,\sin\te\;
\nu_{2}(x)
\\[2mm] \mlab{2.1}
\nu_{\mu}(x) &=&-  \sin\te \;\nu_{1}(x)  \, +\,
\cos\te\;\nu_{2}(x)\; , \eea
Here $\nu_{e}$, $\nu_{\mu}$ are the (Dirac) neutrino
fields with definite flavors. $\nu_{1}$, $\nu_{2}$ are
the (free) neutrino fields with definite masses $m_{1}$,
$m_{2}$, respectively. $\te$ is the mixing angle.
The fields $\nu_{1}$ and $\nu_{2}$ are
expanded as
\be\mlab{2.2}
\nu_{i}(x) = \frac{1}{\sqrt{V}} \sum_{{\bf k},r}
\lf[u^{r}_{{\bf k},i}(t) \al^{r}_{{\bf
k},i}\:+ v^{r}_{-{\bf k},i}(t) \bt^{r\dag }_{-{\bf k},i} \ri]
e^{i {\bf k}\cdot {\bf x}}, \;
\quad i=1,2 \;.
\ee
where $u^{r}_{{\bf k},i}(t)=e^{-i\om_{k,i} t}u^{r}_{{\bf k},i}$
and $v^{r}_{{\bf k},i}(t)=e^{i\om_{k,i} t}v^{r}_{{\bf k},i}$, with
$\om_{k,i}=\sqrt{{\bf k}^2+m_i^2}$. The $\al^{r}_{{\bf k},i}$ and
the $\bt^{r }_{{\bf k},i}$ ($r=1,2$), are the annihilation
operators for the vacuum state
$|0\ran_{1,2}\equiv|0\ran_{1}\otimes |0\ran_{2}$: $\al^{r}_{{\bf
k},i}|0\ran_{1,2}= \bt^{r }_{{\bf k},i}|0\ran_{1,2}=0$. The
anticommutation relations are:
\bea
\mlab{2.3} \{\nu^{\al}_{i}(x),
\nu^{\bt\dag }_{j}(y)\}_{t=t'} = \de^{3}({\bf x}-{\bf y})
\de_{\al\bt} \de_{ij} \;, \;\;\;\;\; \al,\bt=1,..,4 \;,
\\
\mlab{2.4} \{\al^{r}_{{\bf k},i}, \al^{s\dag }_{{\bf q},j}\} =
\de _{{\bf k}q}\de _{rs}\de_{ij}   ;\qquad \{\bt ^{r}_{{\bf k},i},
\bt ^{s\dag}_{{\bf q},j}\} = \de_{{\bf k}q}
\de_{rs}\de_{ij},\;\;\;\; i,j=1,2\;.
\eea
All other anticommutators are zero. The orthonormality and
completeness relations are:
\bea
&&u^{r\dag}_{{\bf k},i} u^{s}_{{\bf k},i} = v^{r\dag}_{{\bf k},i}
v^{s}_{{\bf k},i} = \delta_{rs} \;,\quad  u^{r\dag}_{{\bf k},i}
v^{s}_{-{\bf k},i} = v^{r\dag}_{-{\bf k},i} u^{s}_{{\bf k},i} =
0\;,
\\[2mm]
&&\sum_{r}(u^{r}_{{\bf k},i} u^{r\dag}_{{\bf k},i} +
v^{r}_{-{\bf k},i} v^{r\dag}_{-{\bf k},i}) =\ide \;.
\eea

In QFT the basic dynamics, i.e. the Lagrangian and the resulting
field equations, is given in terms of Heisenberg (or interacting)
fields. The physical observables are expressed in terms of
asymptotic in- (or out-) fields, also called physical or free
fields. In the LSZ formalism of QFT \cite{Itz,Um1}, the free
fields, say for definitiveness the in-fields, are obtained by the
weak limit of the Heisenberg fields for time $t \rar - \infty$.
The meaning of the weak limit is that the realization of the basic
dynamics in terms of the in-fields is not unique so that the limit
for $t \rar -
 \infty$ (or $t \rar + \infty$ for the out-fields) is representation
dependent.

Typical examples are the ones of spontaneously broken symmetry
theories, where the same set of Heisenberg field equations
describes the normal (symmetric) phase as well as the symmetry
broken phase. Since observables are described in terms of
asymptotic fields, unitarily inequivalent representations describe
different, i.e. physically inequivalent, phases. It is therefore
of crucial importance, in order to get physically meaningful
results, to investigate with much care the mapping among
Heisenberg or interacting fields and free fields, i.e. the
dynamical map.

With this warnings, mixing relations such as the relations
(\ref{2.1}) deserve a careful analysis, since they actually
represent a dynamical mapping.  It is now our purpose to
investigate the structure of the Fock spaces ${\cal H}_{1,2}$ and
${\cal H}_{e,\mu}$ relative to $\nu_{1}$, $\nu_{2}$ and
$\nu_{e}$,  $\nu_{\mu}$, respectively. In particular we want
to study the relation among these spaces in the infinite volume
limit.
As usual, we will perform all
computations at finite volume $V$ and only at the end we will put
$V \rar \infty$.

Our first step is the study of the generator of Eqs.(\ref{2.1})
and of the underlying group theoretical structure.
Eqs.(\ref{2.1}) can be recast as \cite{BV95}:
\bea
\nu_{e}^{\al}(x) &=& G^{-1}_{\te}(t)\; \nu_{1}^{\al}(x)\; G_{\te}(t)
\\ [2mm] \mlab{2.8}
\nu_{\mu}^{\al}(x) &=& G^{-1}_{\te}(t)\;
\nu_{2}^{\al}(x)\; G_{\te}(t) \;, \eea
where $G_{\te}(t)$ is given by
\be\mlab{2.9} G_{\te}(t) = \exp\lf[\te \int d^{3}{\bf x}
\lf(\nu_{1}^{\dag}(x) \nu_{2}(x) - \nu_{2}^{\dag}(x) \nu_{1}(x)
\ri)\ri]\;, \ee
and is (at finite volume) an unitary operator:
$G^{-1}_{\te}(t)=G_{-\te}(t)=G^{\dag}_{\te}(t)$, preserving the
canonical anticommutation relations (\ref{2.3}).
Eq.(\ref{2.9}) follows from
$\frac{d^{2}}{d\theta^{2}}\nu^{\alpha}_{e}=-\nu^{\alpha}_{e}\;,\;\;\;
\frac{d^{2}}{d\theta^{2}}\nu^{\alpha}_{\mu}=-\nu^{\alpha}_{\mu}$
with the initial conditions
$\nu^{\alpha}_{e}|_{\theta=0}=\nu^{\alpha}_{1}$,
$\frac{d}{d\theta}\nu^{\alpha}_{e}|_{\theta=0}=\nu^{\alpha}_{2}$
and $\nu^{\alpha}_{\mu}|_{\theta=0}=\nu^{\alpha}_{2}$,
$\frac{d}{d\theta}\nu^{\alpha}_{\mu}|_{\theta=0}=-\nu^{\alpha}_{1}$.

\smallskip

Note that $G_\te$ is an element of  $SU(2)$ since it
can be written as
\bea\mlab{2.11} &&
G_{\te}(t) = \exp[\te(S_{+}(t) - S_{-}(t))]\;,
\\ [2mm]\mlab{2.10}
&&S_{+}(t) =S^{\dag}_{-}(t)\equiv  \int d^{3}{\bf x} \;
\nu_{1}^{\dag}(x) \nu_{2}(x) \,.
\eea
By introducing then
\be\mlab{2.12} S_{3} \equiv \frac{1}{2} \int d^{3}{\bf x}
\lf(\nu_{1}^{\dag}(x)\nu_{1}(x) -
\nu_{2}^{\dag}(x)\nu_{2}(x)\ri)\;, \ee
the $su(2)$ algebra is closed (for $t$ fixed):
\be\mlab{2.14} [S_{+}(t) , S_{-}(t)]=2S_{3} \;\;\;,\;\;\; [S_{3} ,
S_{\pm}(t) ] = \pm S_{\pm}(t) \;. \ee
The action of the mixing generator on the vacuum
$|0 \ran_{1,2}$ is non-trivial and
we have (at finite volume $V$):
\be\mlab{2.22} |0 (t)\ran_{e,\mu}\equiv G^{-1}_{\te}(t)\; |0
\ran_{1,2}\;. \ee
$|0(t) \ran_{e,\mu}$ is the {\em flavor vacuum}, i.e.
the vacuum for the flavor fields.
Note that $G^{-1}_{\te} (t)$ is just
the generator for generalized coherent states of $SU(2)$
\cite{Per}: the flavor vacuum
is therefore an $SU(2)$ (time dependent) coherent state.
Let us
now investigate the infinite volume limit of Eq.(\ref{2.22}).
Using the Gaussian decomposition, $G^{-1}_\te$ is written as
\cite{Per}
\bea\non
\exp[\te(S_{-} - S_{+})]= \exp(-\tan\te \; S_{+})
\;\exp(-2 \ln \: \cos\te \; S_{3}) \;\exp(\tan\te \; S_{-})
\eea
where $0\leq \te < \frac{\pi}{2}$.
We then compute $_{1,2}\lan0|0(t)\ran_{e,\mu}$ and obtain
\bea
{}\hspace{-1cm}
&&_{1,2}\lan0|0(t)\ran_{e,\mu}  = \prod_{{\bf k}}\lf(1- \sin^{2}\te\;|V_{{\bf
k}}|^2\ri)^{2}\equiv \prod_{{\bf k}}\Gamma(k)=e^{\sum_{{\bf
k}}ln\; \Gamma(k)}. \eea
where the function $|V_{{\bf k}}|^2$ is defined in Eq.(\ref{2.39}) and
plotted in Fig.\ref{fig1}. Note that
 $|V_{\bf k}|^2$ depends on ${|\bf k}|$, it is
always in the interval $[0,\frac{1}{2}[$ and goes to zero for
$|{\bf k}| \rar \infty$. By using the
customary continuous limit relation $\sum_{{\bf k}}\;\rar \;
\frac{V}{(2\pi)^{3}}\int d^{3}{\bf k}$, in the infinite volume
limit we obtain (for any $t$)
\be\mlab{2.34} \lim_{V \rar \infty}\; _{1,2}\lan0|0(t)\ran_{e,\mu} =
\lim_{V \rar \infty}\; e^{\frac{V}{(2\pi)^{3}}\int d^{3}{\bf k}
\;ln\; \Gamma(k)}= 0 \ee
since $\Gamma({\bf k}) < 1$ for any value of ${\bf k}$ and of  $m_{1}$ and $m_{2}$ (with
 $m_{2}\neq m_1$).

Notice that (\ref{2.34}) shows that the orthogonality between
$|0(t)\ran_{e,\mu}$ and $|0\ran_{1,2}$ is due to the infrared
contributions which are taken in care by the infinite volume limit
and therefore high momentum contributions do not influence the
result (for this reason here we do not need to consider the
regularization problem of the UV divergence of the integral of
$\ln\;\Ga({\bf k})$). Of course, this orthogonality disappears
when $\te =0$ and/or when $m_{1} = m_{2}$ (in this case
$V_{{\bf k}}=0$ for any ${\bf k}$ ).

Eq.(\ref{2.34}) expresses the unitary inequivalence  in the
infinite volume limit of the flavor and the mass representations
and shows the non-trivial nature of the mixing transformations
(\ref{2.1}), resulting in the condensate structure of the
flavor vacuum. In
Section \ref{oscill} we will see how such a vacuum structure leads to
phenomenological consequences in the neutrino oscillations, which
may be possibly  experimentally tested.

By use of $G_{\te}(t)$, the flavor fields can be expanded as:
\bea\label{exnue1}
&&{}\hspace{-1.5cm}
\nu_\si(x)\,=\, \sum_{r=1,2} \int \frac{d^3
\bk}{(2\pi)^\frac{3}{2}} \lf[ u^{r}_{{\bf k},i}(t) \al^{r}_{{\bf k},\si}(t)
+    v^{r}_{-{\bf k},i}(t) \bt^{r\dag}_{-{\bf k},\si}(t) \ri]
e^{i {\bf k}\cdot{\bf x}}\,, \eea
with $(\si,i)=(e,1), (\mu,2)$.
The flavor annihilation operators are defined as $\al^{r}_{{\bf
k},\si}(t) \equiv G^{-1}_{\bf \te}(t)\al^{r}_{{\bf k},i} G_{\bf
\te}(t)$ and $\bt^{r\dag}_{{-\bf k},\si}(t)\equiv
 G^{-1}_{\bf \te}(t) \bt^{r\dag}_{{-\bf k},i}
G_{\bf \te}(t)$.
In  the reference frame such
that ${\bf k}=(0,0,|{\bf k}|)$, we have the simple expressions:
\bea
&&{}\hspace{-1cm}
\al^{r}_{{\bf k},e}(t)=\cos\te\;\al^{r}_{{\bf
k},1}\;+\;\sin\te\;\lf( U_{{\bf k}}^{*}(t)\; \al^{r}_{{\bf
k},2}\;+\;\ep^{r}\;
V_{{\bf k}}(t)\; \bt^{r\dag}_{-{\bf k},2}\ri)
\\ \mlab{2.36}
&&{}\hspace{-1cm}
\al^{r}_{{\bf k},\mu}(t)=\cos\te\;\al^{r}_{{\bf
k},2}\;-\;\sin\te\;\lf( U_{{\bf k}}(t)\; \al^{r}_{{\bf
k},1}\;-\;\ep^{r}\;
V_{{\bf k}}(t)\; \bt^{r\dag}_{-{\bf k},1}\ri)
\\
&&{}\hspace{-1.2cm}
\bt^{r}_{-{\bf k},e}(t)=\cos\te\;\bt^{r}_{-{\bf
k},1}\;+\;\sin\te\;\lf( U_{{\bf k}}^{*}(t)\; \bt^{r}_{-{\bf
k},2}\;-\;\ep^{r}\;
V_{{\bf k}}(t)\; \al^{r\dag}_{{\bf k},2}\ri)
\\
&&{}\hspace{-1.2cm}\bt^{r}_{-{\bf k},\mu}(t)=\cos\te\;\bt^{r}_{-{\bf
k},2}\;-\;\sin\te\;\lf( U_{{\bf k}}(t)\; \bt^{r}_{-{\bf
k},1}\;+\;\ep^{r}\; V_{{\bf k}}(t)\; \al^{r\dag}_{{\bf k},1}\ri)
\eea
where $\ep^{r}=(-1)^{r}$ and
\bea &&{}\hspace{-1.2cm}
 U_{{\bf k}}(t)\equiv u^{r\dag}_{{\bf
k},2}(t)u^{r}_{{\bf k},1}(t)= v^{r\dag}_{-{\bf
k},1}(t)v^{r}_{-{\bf k},2}(t) =\, |U_{{\bf
k}}|\,e^{i(\om_{k,2}-\om_{k,1})t}
\\ [2mm] \mlab{2.37} &&
{}\hspace{-1cm}V_{{\bf k}}(t)\equiv \ep^{r}\;u^{r\dag}_{{\bf
k},1}(t)v^{r}_{-{\bf k},2}(t)= -\ep^{r}\;u^{r\dag}_{{\bf
k},2}(t)v^{r}_{-{\bf k},1}(t) = \,|V_{{\bf
k}}|\;e^{i(\om_{k,2}+\om_{k,1})t}
\\ [2mm] \mlab{2.38}
&&|U_{{\bf k}}|=\frac{|{\bf k}|^{2} +(\om_{k,1}+m_{1})(\om_{k,2}+m_{2})}{2
\sqrt{\om_{k,1}\om_{k,2}(\om_{k,1}+m_{1})(\om_{k,2}+m_{2})}}
\\ \mlab{2.39} &&
|V_{{\bf k}}|=\frac{ (\om_{k,1}+m_{1}) - (\om_{k,2}+m_{2})}{2
\sqrt{\om_{k,1}\om_{k,2}(\om_{k,1}+m_{1})(\om_{k,2}+m_{2})}}\, |{\bf k}|
\\ \mlab{2.40}
&&|U_{{\bf k}}|^{2}+|V_{{\bf k}}|^{2}=1.
\eea
%
\begin{figure}[t]
\setlength{\unitlength}{1mm} \vspace*{70mm} 
\includegraphics{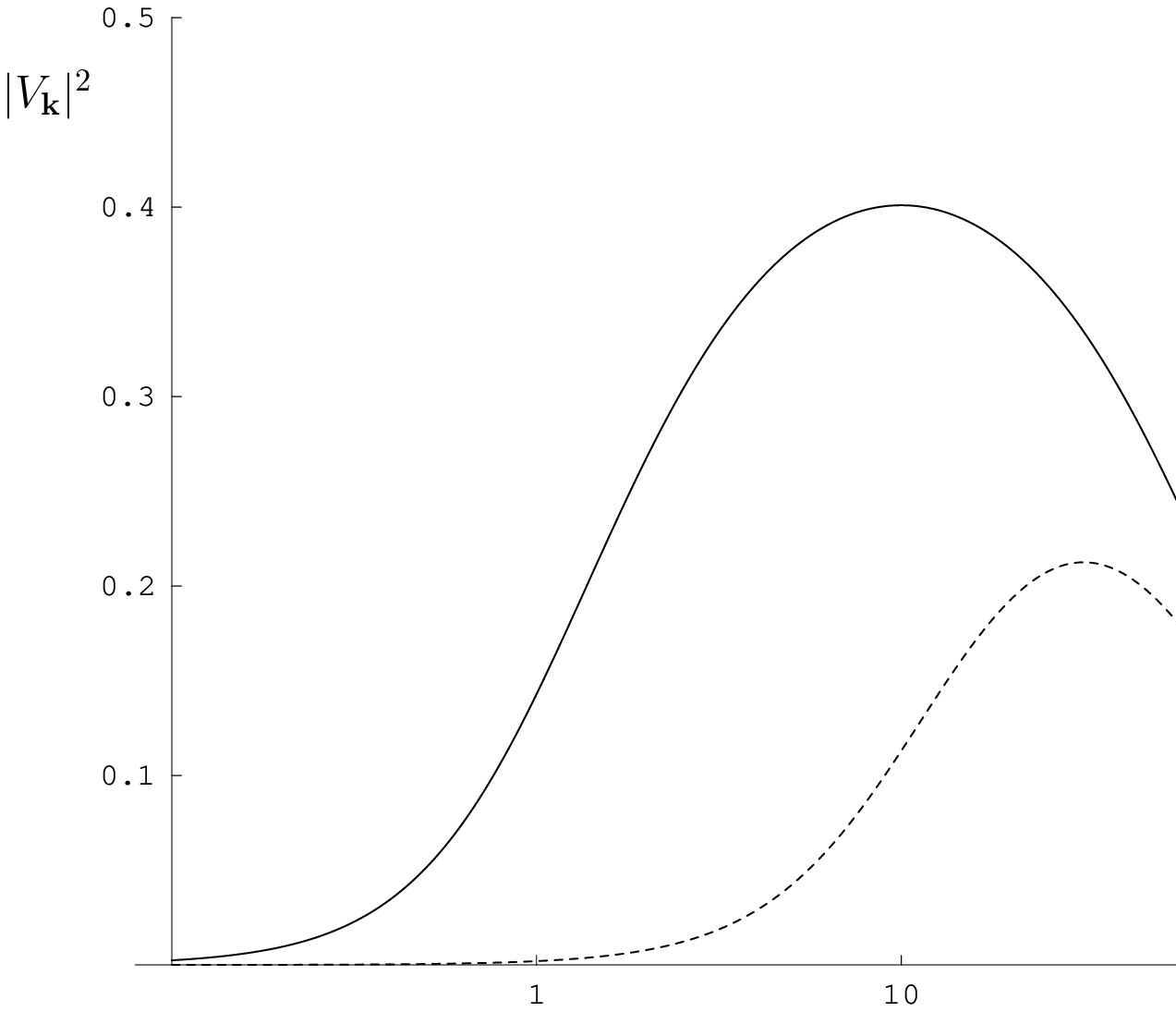}
\caption{The fermion condensation density $|V_{\bf k}|^2$ as a function
of  $|{\bf k}|$ for $m_1=1$, $m_2=100$ (solid line)
and $m_1=10$, $m_2=100$ (dashed line).}

\mlab{fig1}
\end{figure}
%
%
%
The condensation density of the flavor vacuum is given by
\bea\mlab{2.41b} &&\;_{e,\mu}\lan 0(t)| \al_{{\bf k},i}^{r \dag}
\al^r_{{\bf k},i} |0(t)\ran_{e,\mu}=  \sin^{2}\te\; |V_{{\bf k}}|^{2}
, \qquad i=1,2\,,
\eea
with the same result for antiparticles\footnote{In the case of
three flavors \cite{BV95,3flavors}, the condensation densities
are different for different $i$ and for antiparticles (when CP violation is
present)}.
Note that the $|V_{{\bf k}}|^2$ has a maximum at $\sqrt{m_1 m_2}$
and $|V_{{\bf k}}|^2\simeq \frac{(m_\2 -m_\1)^2}{4 |{\bf k}|^2}$
 for $ |{\bf k}|\gg\sqrt{m_\1 m_\2}$.

\subsection{Boson mixing}
\label{bosonmix}

Let us now consider boson mixing \cite{lathuile,bosonmix}
in the case of charged fields.
We define the mixing relations as:
\bea\non
\phi_{A}(x) &=&\cos\te \; \phi_{1}(x)   +  \sin\te \; \phi_{2}(x)
\\[2mm] \label{bosmix}
\phi_{B}(x) &=&-  \sin\te \; \phi_{1}(x)  +  \cos\te \; \phi_{2}(x)
\eea
where generically we denote the mixed fields with suffixes $A$ and $B$.
Let the fields $\phi_{i}(x)$, $i=1,2$, be free complex fields with
definite masses. Their
conjugate momenta are $\pi_{i}(x)=\pa_{0}\phi_{i}^{\dag}(x)$ and
the commutation relations are the usual ones:
\bea \lf[\phi_{i}(x),\pi_{j}(y)\ri]_{t=t'}=
\lf[\phi_{i}^{\dag}(x),\pi_{j}^{\dag}(y)\ri]_{t=t'}=i\de^{3}
({\bf x}-{\bf y}) \, \de_{ij} \lab{2.50} \eea
with $i,j=1,2$ and the other equal--time commutators vanishing.
The Fourier expansions of fields and momenta are:
\bea\lab{2.51} \phi_{i}(x) = \int \frac{d^3 {\bf k}}{(2\pi)^{\frac{3}{2}}}
\frac{1}{\sqrt{2\om_{k,i}}} \lf( a_{{\bf k},i}\, e^{-i \om_{k,i} t} +
b^{\dag }_{-{\bf k},i}\, e^{i \om_{k,i} t}  \ri) e^{i {\bf k}\cdot {\bf
x}}
\\ [2mm]\lab{2.52}
\pi_{i}(x) = i\,\int \frac{d^3 {\bf k}}{(2\pi)^{\frac{3}{2}}}
\sqrt{\frac{\om_{k,i}}{2}} \lf( a^{\dag }_{{\bf k},i}\, e^{i \om_{k,i}
t} - b_{-{\bf k},i}\, e^{-i \om_{k,i} t} \ri) e^{i {\bf k}\cdot {\bf
x}}\,, \eea
where $\om_{k,i}=\sqrt{{\bf k}^2
+ m_i^2}$ and $\lf[a_{{\bf k},i},a_{{\bf p},j}^{\dag} \ri]= \lf[b_{{\bf
k},i},b_{{\bf p},j}^{\dag} \ri]=\de^{3}({\bf k}-{\bf p}) \de_{ij}\, ,$
with $i,j=1,2\,$ and the other commutators vanishing.

We proceed in a similar way as for fermions and write
Eqs.(\ref{bosmix}) as
\bea\lab{2.53b}
\phi_{\si}(x) &=& G^{-1}_\te(t)\; \phi_{i}(x)\; G_\te(t)
\eea
with \small $(\si,i)=(A,1),(B,2)$, \normalsize
and similar expressions for $\pi_{A}$, $\pi_B$. We have
\bea\lab{2.55}
G_\te(t) = \exp[\te(S_{+}(t) - S_{-}(t))] ~.
\eea
The operators
\bea\lab{2.55b}
&&S_+(t)= S_{-}^{\dag}(t) \equiv -i\;\int d^3{\bf x} \;
(\pi_{1}(x)\phi_{2}(x) - \phi_{1}^{\dag}(x)\pi_{2}^{\dag}(x)) \,,
\\[2mm] \lab{2.56}
&&{}\hspace{-.6cm} \non
S_{3} \equiv \frac{-i}{2} \int d^3 {\bf x}
\lf(\pi_{1}(x)\phi_{1}(x) - \phi_{1}^{\dag}(x)\pi_{1}^{\dag}(x)
- \pi_{2}(x)\phi_{2}(x)
+\phi_{2}^{\dag}(x) \pi_{2}^{\dag}(x)
\ri)
\\ &&
\eea
close the $su(2)$ algebra (at a given $t$).

As for fermions, the action of the generator of the
mixing transformations on the vacuum $|0 \ran_{1,2}$ for the
fields $\phi_{1,2}$ is non-trivial and
induces on it a $SU(2)$ coherent state structure \cite{Per}:
\bea\label{2.61} |0(t) \ran_\AB \equiv G^{-1}_\te(t)\; |0
\ran_{1,2}\,. \eea
We will refer to the state $|0(t) \ran_\AB$ as to
the flavor vacuum for bosons.
The orthogonality between $|0(t) \ran_\AB$
and $|0 \ran_{1,2}$ can be proved \cite{bosonmix}.
The Fourier expansion for the flavor fields is:
\bea\lab{flavAB} \phi_{\si}(x) = \int \frac{d^3 {\bf k}}{(2\pi)^{\frac{3}{2}}}
\frac{1}{\sqrt{2\om_{k,i}}} \lf( a_{{\bf k},\si}(t)\, e^{-i \om_{k,i} t} +
b^{\dag }_{-{\bf k},\si}(t)\, e^{i \om_{k,i} t}  \ri) e^{i {\bf k}\cdot {\bf
x}}
\eea
with \small $(\si,i)=(A,1),(B,2)$, \normalsize
and similar expressions for $\pi_{A}$, $\pi_B$.

The annihilation operators  for the vacuum $|0(t) \ran_\AB$
are defined  $a_{{\bf k},A}(t) \equiv G^{-1}_\te(t) \; a_{{\bf
k},1}\;G_\te(t)$, etc. We have:
\bea \label{2.62a}
&&{}\hspace{-1cm}
a_{{\bf k},A}(t)\,=\,\cos\te\;a_{{\bf k},1}\;+\;\sin\te\;\lf(
U^*_{{\bf k}}(t)\; a_{{\bf k},2}\;+\; V_{{\bf k}}(t)\;
b^{\dag}_{-{\bf k},2}\ri)\, ,
\\
&&{}\hspace{-1cm}
a_{{\bf k},B}(t)\,=\,\cos\te\;a_{{\bf k},2}\;-\;\sin\te\;\lf(
U_{{\bf k}}(t)\; a_{{\bf k},1}\;-
\; V_{{\bf k}}(t)\; b^{\dag}_{-{\bf k},1}\ri)\, ,
\\
&&{}\hspace{-1cm}
b_{-{\bf k},A}(t)\,=\,\cos\te\;b_{-{\bf k},1}\;+\;\sin\te\;\lf(
U^*_{{\bf k}}(t)\; b_{-{\bf k},2}\;+
\; V_{{\bf k}}(t)\; a^{\dag}_{{\bf k},2}\ri)\, ,
\\ \label{2.62d}
&&{}\hspace{-1cm}
b_{-{\bf k},B}(t)\,=\,\cos\te\;b_{-{\bf k},2}\;-\;\sin\te\;\lf(
U_{{\bf k}}(t)\; b_{-{\bf k},1}\;-
\; V_{{\bf k}}(t)\; a^{\dag}_{{\bf k},1}\ri) ~.
\eea
These operators satisfy the canonical commutation relations (at equal
times).
%
\begin{figure}[t]
\setlength{\unitlength}{1mm} \vspace*{70mm} 
\includegraphics{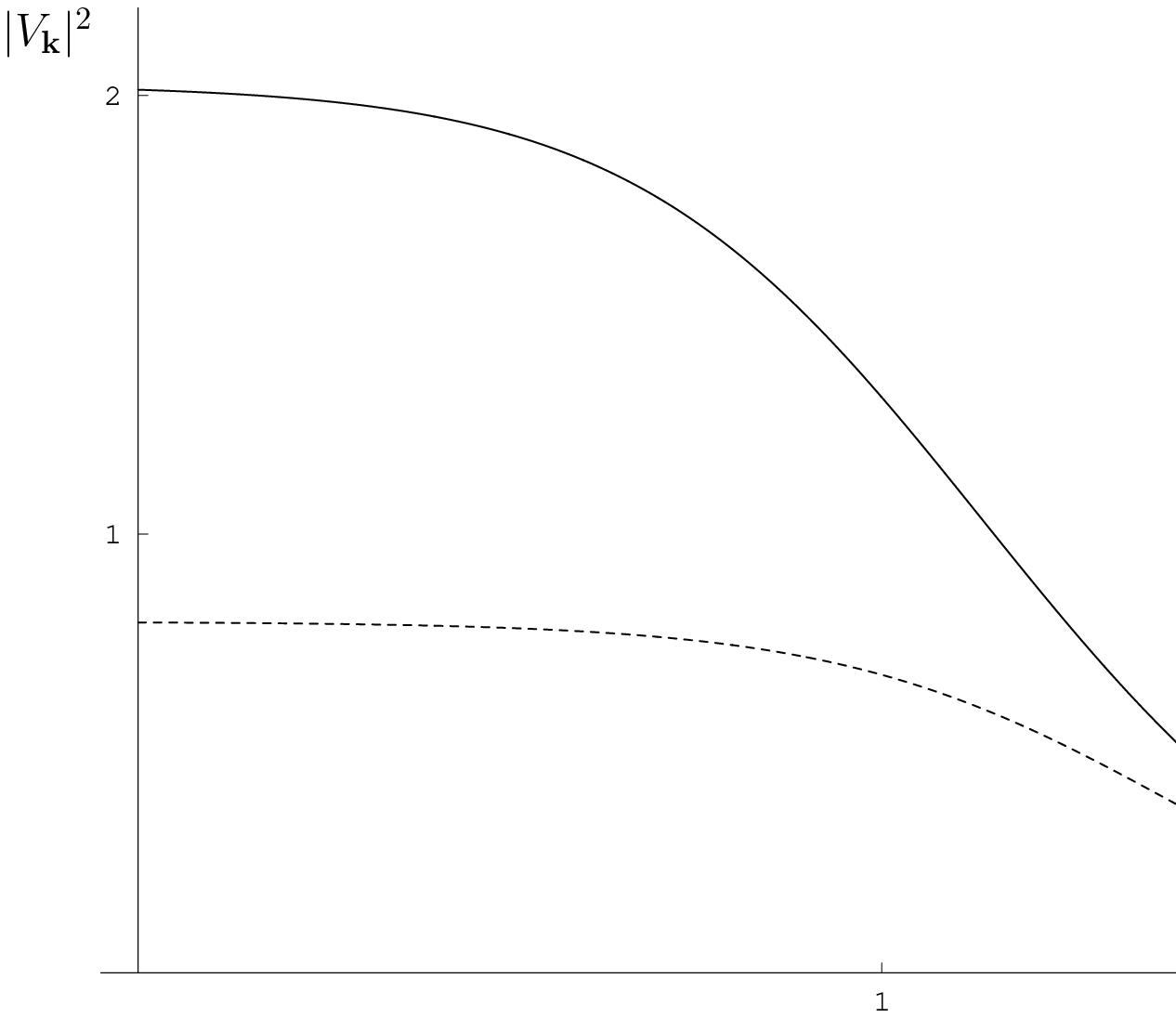}
\caption{The boson condensation density
$|V_{\bf k}|^2$ as a function of $|{\bf k}|$
for $m_1=1$, $m_2=10$  (solid line) and $m_1=2$, $m_2=10$ (dashed line).}

\mlab{fig2}
\end{figure}
%
As for the case of the fermion mixing, the structure of  the flavor
ladder operators Eqs.(\ref{2.62a})-(\ref{2.62d})
is recognized to be the one of a rotation
combined with a
Bogoliubov transformation.
Indeed, in the above equations appear the
Bogoliubov coefficients:
\bea \lab{bog1}
&&{}\hspace{-1cm}
U_{{\bf k}}(t)\equiv |U_{{\bf k}}| \; e^{i(\om_{k,2}-
\om_{k,1})t} \quad , \quad V_{{\bf k}}(t)\equiv |V_{{\bf k}}| \;
e^{i(\om_{k,1}+ \om_{k,2})t}
\\ [2mm]  \lab{bog2}
&&{}\hspace{-1cm}
|U_{{\bf k}}|\equiv \frac{1}{2}
\lf( \sqrt{\frac{\om_{k,1}}{\om_{k,2}}}
+ \sqrt{\frac{\om_{k,2}}{\om_{k,1}}}\ri) \;, \quad
|V_{{\bf k}}|\equiv  \frac{1}{2} \lf(
\sqrt{\frac{\om_{k,1}}{\om_{k,2}}}
- \sqrt{\frac{\om_{k,2}}{\om_{k,1}}} \ri)
\\ [2mm]  \lab{bog3}
&&{}\hspace{-1cm}
|U_{{\bf k}}|^{2}-|V_{{\bf k}}|^{2}=1\,,
\eea
Note the difference with respect to the fermionic case Eq.(\ref{2.40}).

The condensation density of the flavor vacuum is given for any $t$ by
\bea\label{2.63} {}_\AB\lan 0(t)| a_{{\bf k},i}^{\dag} a_{{\bf k},i}
|0(t)\ran_\AB=\,
\sin^{2}\te\;  |V_{{\bf k}}|^{2}, \qquad i=1,2\,,
\eea
with same result for antiparticles.
The function $|V_{{\bf k}}|^{2}$ is maximal at $|{\bf k}|=0$
($|V_{max}|^2 = \frac{(m_1 -m_2)^2}{4 m_1 m_2}$) and $|V_{{\bf k}}|^2\simeq
\lf(\frac{\De m^2}{4 |{\bf k}|^2}\ri)^2 $
for $|{\bf k}|^2 \gg\frac{m_\1^2 +m_\2^2}{2}$.
A plot is given in Fig.\ref{fig2} for sample values of the masses.

\subsection{Currents and charges for mixed fields}
\label{currents}

We now study the transformations acting on a doublet of free
fields with different masses. The results of this Section clarify
the meaning of the $su(2)$ algebraic structure found before and
will be useful in the discussion of neutrino oscillations.

\subsubsection{Fermions}
\label{fermcurr}

Let  us consider the Lagrangian for two free
Dirac fields, with masses $m_1$ and $m_2$:
\bea\label{lagf12}
{\cal L}(x)\,=\,  {\bar \Psi_m}(x) \lf( i \not\!\partial -
  M_d\ri) \Psi_m(x)
\eea
where $\Psi_m^T=(\nu_1,\nu_2)$ and $M_d = diag(m_1,m_2)$. We
introduce a subscript $m$ to denote quantities which are in terms of
fields with definite masses.

 ${\cal L}$ is invariant under global
$U(1)$ phase transformations  of the type $\Psi_m' \, =\, e^{i \al
}\, \Psi_m$: as a result, we have the conservation of the Noether
charge $Q=\int d^3{\bf x} \, I^0(x) $ (with $I^\mu(x)={\bar \Psi}_m(x)
\, \ga^\mu \, \Psi_m(x)$)  which is indeed the total charge of the
system (i.e. the total lepton number).
Consider then the global $SU(2)$ transformation \cite{currents}:
\bea
\Psi_m'(x) \, =\, e^{i \al_j \tau_j }\, \Psi_m (x),\qquad
\qquad j \,=\, 1, 2, 3. \eea
with $\tau_j=\si_j/2$  and $\si_j$ being
the Pauli matrices.
For $m_1 \neq m_2$, the Lagrangian is
not generally invariant under the above transformations. We have indeed:
\bea\de {\cal L}(x)&= &  i \al_j \,{\bar \Psi_m}(x)\, [\tau_j,M_d ]\,
\Psi_m (x)\, =\,  - \al_j \,\pa_\mu J_{m,j}^\mu(x)
\\[2mm]
J^\mu_{m,j}(x)& =&  {\bar \Psi_m}(x)\, \ga^\mu\, \tau_j\, \Psi_m(x),\qquad
\qquad j \,=\, 1, 2, 3.
\eea
Explicitly:
\bea
J^\mu_{m,1}(x) &=&\frac{1}{2} \lf[{\bar \nu}_\1(x) \, \ga^\mu\,\nu_\2(x)
\,+\, {\bar \nu}_\2(x) \, \ga^\mu\,\nu_\1(x) \ri]
\\ [2mm]
J^\mu_{m,2}(x) &=&\frac{i}{2} \lf[ {\bar \nu}_\1(x) \, \ga^\mu\,\nu_\2(x) \,-\,
{\bar \nu}_\2(x)\, \ga^\mu\, \nu_\1(x) \ri]
\\ [2mm]
J^\mu_{m,3}(x) &=&\frac{1}{2} \lf[ {\bar \nu}_\1(x)\, \ga^\mu\, \nu_\1(x) \,-\,
{\bar \nu}_\2(x)\, \ga^\mu\, \nu_\2(x)\ri]
\eea
The  charges $Q_{m,j}(t)\equiv \int d^3 {\bf x} \,J^0_{m,j}(x)$
satisfy the $su(2)$ algebra (at equal times):
$[Q_{m,j}(t),Q_{m,k}(t)]\, =\, i \,\ep_{jkl} \, Q_{m,l}(t)\, $.
Note that $2 Q_{m,2}(t)$ is indeed the generator of mixing
transformations introduced in \S\ref{fermionmix}. Also note that Casimir
operator is proportional to the total (conserved) charge: $C_{m}
\,=\,\frac{1}{2}Q$ and that, since $Q_{m,3}$  is conserved in
time, we have
\bea\label{noether1}
&&{}\hspace{-1cm}
Q_{1}\, \equiv \,\frac{1}{2}Q \,+ \,Q_{m,3}
\quad, \quad Q_{2}\, \equiv \,\frac{1}{2}Q \,- \,Q_{m,3}
\\
&&{}\hspace{-1cm}
Q_i \, = \,\sum_{r} \int d^3 {\bf k}\lf( \al^{r\dag}_{{\bf k},i}
\al^{r}_{{\bf k},i}\, -\, \bt^{r\dag}_{-{\bf k},i}\bt^{r}_{-{\bf
k},i}\ri) \quad, \quad i=1, 2.
\eea
These are nothing but  the Noether charges associated with the
non-interacting fields $\nu_1$  and $\nu_2$: in the absence of
mixing, they are the flavor charges,  separately
conserved for each generation.

\subsubsection{Bosons}
\label{boscurr}

The above analysis can be easily extended to the boson case.
We consider the Lagrangian
\be\lab{lagb12} {\cal L}(x)\,=\, \pa_\mu \Phi_m^\dag(x) \pa^\mu \Phi_m(x) \,
- \, \Phi_m^\dag(x) M_d  \Phi_m(x) \ee
with $\Phi_m^T=(\phi_1,\phi_2)$ being charged scalar fields
and $M_d = diag(m_1^2,m_2^2)$.

We have now \cite{currents}
\bea\lab{rotgen} %
\Phi_m'(x) & =& e^{i \al_j \, \tau_j}\, \Phi_m(x)
\\ [2mm]
\de {\cal L}(x)&= &  i \,\al_j\, \Phi_m^\dag(x)
\, [\tau_j\,, \,M_d] \, \Phi_m(x) \, = \, -\al_j \,\pa_\mu\,
J^\mu_{m,j}(x)\, ,
\\ [2mm]
J^\mu_{m,j}(x) &=& i\, \Phi_m^\dag(x) \, \tau_j\,
\stackrel{\lrar}{\pa^\mu}\, \Phi_m(x)\; , \qquad j \, =\, 1,2,3.
\eea
Again, the corresponding charges $Q_{m,j}(t)$ satisfy the $su(2)$
algebra and the mixing generator for bosons is proportional to
$Q_{m,2}(t)$.

\subsection[Generalization of mixing
transformations]{Generalization of mixing transformations}
\label{general}

We have seen in \S\ref{fermionmix} how the fields $\nu_e$ and $\nu_\mu$ can be
expanded in the same bases as $\nu_1$ and $\nu_2$,
see Eq.(\ref{exnue1}).
As observed in Ref.\cite{fujii1}, however, such a choice
is actually a special one, and a more
general possibility exists.
Indeed, in  the expansion Eq.(\ref{exnue1}) one could
use eigenfunctions with arbitrary masses $\mu_\sigma$  and write
the flavor fields as \cite{fujii1}:
\begin{eqnarray}\label{exnuf2}
{}\hspace{-1cm}
&& \nu_{\sigma}(x) \,=\, \sum_{r=1,2} \int \frac{d^3
\bk}{(2\pi)^\frac{3}{2}}  \lf[
u^{r}_{{\bf k},\sigma} {\widetilde \alpha}^{r}_{{\bf k},\sigma}(t)
+ v^{r}_{-{\bf k},\sigma} {\widetilde \beta}^{r\dag}_{-{\bf
k},\sigma}(t) \ri]  e^{i {\bf k}\cdot{\bf x}} ,
\end{eqnarray}
where $u_{\sigma}$ and $v_{\sigma}$ are the
eigenfunctions with mass $\mu_\sigma$ ($\si=e,\mu$).
We denote by a tilde the
generalized flavor operators introduced in Ref.\cite{fujii1}.
The expansion Eq.(\ref{exnuf2}) is more
general than the one in Eq.(\ref{exnue1}) since the latter
corresponds to the particular choice $\mu_e\equiv m_1$, $\mu_\mu
\equiv m_2$.
The relation between the general flavor operators of
Eq.(\ref{exnuf2}) and those of Eq.(\ref{exnue1}) is
\begin{eqnarray}\label{FHYBVa}
&&\lf(\begin{array}{c} {\widetilde \alpha}^{r}_{{\bf k},\sigma}(t)\\
{\widetilde \beta}^{r\dag}_{{-\bf k},\sigma}(t)
\end{array}\ri)
= J^{-1}(t)  \lf(\begin{array}{c} \alpha^{r}_{{\bf
k},\sigma}(t)\\ \beta^{r\dag}_{{-\bf k},\sigma}(t)
\end{array}\ri)J(t) ~~,
\\ \label{FHYBVb}\non
&&{}\hspace{-1cm}
J(t)= \prod_{{\bf k}, r}\, \exp\lf\{ i
\mathop{\sum_{(\sigma,j)}} \xi_{\sigma,j}^{\bf k}\lf[
\alpha^{r\dag}_{{\bf k},\sigma}(t)\beta^{r\dag}_{{-\bf
k},\sigma}(t) + \beta^{r}_{{-\bf k},\sigma}(t)\alpha^{r}_{{\bf
k},\sigma}(t) \ri]\ri\}\,.
\end{eqnarray}
where $\xi_{\si,j}^{\bf k}\equiv (\chi_\si - \chi_j)/2$
with $\cot\chi_{\si} = |{\bf k}|/\mu_{\si}$ and
$\cot\chi_{j} = |{\bf k}|/m_j$.

Thus the Hilbert space for the flavor fields
is not unique: an infinite number of vacua
can be generated by introducing
the arbitrary mass parameters $\mu_\si$. It is obvious
that physical quantities must not depend on these parameters.
Similar results are valid for bosons, see Ref.\cite{bosonmix}.

\section{Flavor oscillations in QFT}
\label{oscill}

As an application of the theoretical scheme above
developed, we study flavor oscillations, both for fermions
and for bosons. The QFT  treatment  leads to
exact oscillation formulas exhibiting corrections with respect to
the usual QM ones.

\subsection{Neutrino oscillations}
\label{neuosc}

Let  us now return to the  Lagrangian Eq.(\ref{lagf12}) and write
it in the flavor basis
(subscript $f$ denotes here flavor)
\bea \label{lagfem}
{\cal L}(x)\,=\,  {\bar \Psi_f}(x) \lf( i \not\!\partial -
  M \ri) \Psi_f(x)
\eea
where $\Psi_f^T=(\nu_e,\nu_\mu)$ and $M = \lf(\ba{cc} m_e &
m_{e\mu} \\ m_{e\mu} & m_\mu \ea \ri)$. Obviously, ${\cal L}$
is still
invariant under $U(1)$. We then consider the
$SU(2)$ transformation \cite{currents}:
\bea
\Psi_f'(x) &=& e^{i \al_j \tau_j }\, \Psi_f(x),
\\[2mm]
\de {\cal L}(x) &=&   i \al_j\,{\bar \Psi_f}(x)\,
[\tau_j, M]\, \Psi_f(x)\, =\, - \al_j \,\pa_\mu J_{f,j}^{\mu}(x)\,,
\\ [2mm]
J^\mu_{f,j}(x)&=&  {\bar \Psi_f}(x)\, \ga^\mu\, \tau_j\,
\Psi_f(x),\quad \qquad  j \,=\, 1, 2, 3.
\eea
The charges $Q_{f,j}(t)\equiv \int d^3{\bf x} \,J^0_{f,j}(x)$
 satisfy the $su(2)$ algebra. Note that, because of
the off--diagonal (mixing) terms in the mass matrix $M$,
$Q_{f,3}$ is not anymore conserved. This implies an exchange of
charge between $\nu_e$ and $\nu_\mu$, resulting in the phenomenon
of flavor oscillations.

Let us indeed define the {\em flavor charges} for mixed fields as
\bea
Q_e(t) & \equiv & \int d^3 {\bf x}\,\nu_e^\dag(x)\nu_e(x) \,=\,
\frac{1}{2} Q \, + \, Q_{f,3}(t)
\\
Q_\mu(t) & \equiv & \int d^3 {\bf x}\,\nu_\mu^\dag(x)\nu_\mu(x) \,=\,
\frac{1}{2} Q \, -  \, Q_{f,3}(t)
\eea
where $Q_e(t) \, + \,Q_\mu(t) \, = \, Q$.
They are related
to the Noether charges  as
\bea\label{chaconn}
Q_\si(t) &=& G_\te^{-1}(t)\,Q_i \, G_\te(t)
\eea
with ${\small (\si,i)=(e,1),(\mu,2)}$.
From Eq.(\ref{chaconn}), it follows that the flavor charges
are diagonal in the flavor ladder operators:
\bea
{}\hspace{-.8cm}
Q_\sigma(t) & = & \sum_{r}\int d^3 {\bf k} \lf( \alpha^{r\dag}_{{\bf
k},\sigma}(t) \alpha^{r}_{{\bf k},\sigma}(t)\, -\,
\beta^{r\dag}_{-{\bf k},\sigma}(t)\beta^{r}_{-{\bf
k},\sigma}(t)\ri)\,,
\eea
with $\sigma= e,\mu$.
We work in the Heisenberg picture and
 define the state for a particle with definite (electron) flavor,
spin and momentum as\footnote{Similar results are obtained for
a muon neutrino state: $|\alpha^r_{{\bf k},\mu}\rangle \equiv
\alpha^{r \dagger}_{{\bf k},\mu}(0)|0\rangle_{e,\mu}$.}:
\be
|\alpha^r_{{\bf k},e}\rangle \equiv
\alpha^{r \dagger}_{{\bf k},e}(0)|0\rangle_{e,\mu} =
G_\theta^{-1}(0) \alpha^{r \dagger}_{{\bf k},1}|0\rangle_\mass\;,
\ee
where $|0\rangle_{e,\mu}\equiv |0(0)\rangle_{e,\mu}$.
Note that the $|\alpha^r_{{\bf k},e}\rangle $ is an eigenstate of
$Q_e(t)$, at $t=0$: $Q_e(0)|\alpha^r_{{\bf k},e}\rangle =
|\alpha^r_{{\bf k},e}\rangle $.
We thus have $ \;_{e,\mu}\langle 0|Q_\si(t)| 0\rangle_{e,\mu}\;
 =\; 0$ and
\bea\non
{\cal Q}_{{\bf k},\si}(t) &\equiv&
\langle \alpha^r_{{\bf k},e}|
Q_\sigma(t) |\alpha^r_{{\bf k},e}\rangle
\\ [2mm] \label{charge2}
&= &\lf|\lf \{\al^{r}_{{\bf k},\si}(t), \al^{r
\dag}_{{\bf k},\rho}(0) \ri\}\ri|^{2} \;+ \;\lf|\lf\{\bt_{{-\bf
k},\si}^{r \dag}(t), \al^{r \dag}_{{\bf k},\rho}(0) \ri\}\ri|^{2}
\eea
Charge conservation is ensured at any time: ${\cal
Q}_{{\bf k},e}(t) + {\cal Q}_{{\bf k},\mu}(t)\; = \; 1$. The oscillation formulas for
the flavor charges are then \cite{BHV99}
\bea\non
{}\hspace{-1cm}
{\cal Q}_{{\bf k},e}(t)&=& 1 \,-\, \sin^{2}( 2 \theta)\, |U_{{\bf k}}|^{2} \; \sin^{2}
\lf( \frac{\omega_{k,2} - \omega_{k,1}}{2} t \ri)
\\ [2mm] \label{enumber}
&&+  \sin^{2}( 2 \theta)\, |V_{{\bf
k}}|^{2} \; \sin^{2} \lf( \frac{\omega_{k,2} + \omega_{k,1}}{2}
t \ri)  \, ,
\\[5mm] \non
{}\hspace{-1cm}
{\cal Q}_{{\bf k},\mu}(t)&=&  \sin^{2}( 2 \theta)\, |U_{{\bf k}}|^{2} \; \sin^{2}
\lf( \frac{\omega_{k,2} - \omega_{k,1}}{2} t \ri)
\\ [2mm]\label{munumber}
&&+\sin^{2}( 2 \theta)\,
|V_{{\bf k}}|^{2} \; \sin^{2} \lf( \frac{\omega_{k,2} + \omega_{k,1}}{2}
t \ri) \, .
\end{eqnarray}
This result is exact. There are two differences with respect to
the usual formula for neutrino oscillations: the amplitudes are
energy dependent, and there is an additional oscillating term.

In the relativistic limit ($|{\bf k}|\gg \sqrt{m_\1m_\2}$)  we obtain ($\te=\pi/4$):
\bea\non
&&{}\hspace{-1.8cm}
{\cal Q}_{{\bf k},\mu}(t)
\simeq    \, \lf( 1- \frac{(\De m)^2}{4 |{\bf k}|^2}\ri)
\sin^2 \lf[ \frac{\De m^2}{4 |{\bf k}|} \, t \ri]
\,
\\
&&{}\hspace{2cm}+  \, \frac{(\De m)^2}{4 k^2} \sin^2 \lf[\lf(|{\bf k}| +
\frac{m_\1^2+m_\2^2}{4|{\bf k}|}\ri) \, t \ri]\,.
\eea
The usual QM formulas \cite{Pont},
 are thus approximately recovered.
Observe that for small times  we have:
\bea
{}\hspace{-1.2cm}
{\cal Q}_{{\bf k},\mu}(t)
\simeq    \frac{(m_\2-m_\1)^2}{4} \, \lf( 1\,+\,
\frac{m_\1^2+m_\2^2}{2 |{\bf k}|^2}\,+\,
\frac{(m_\1+m_\2)^2}{4 |{\bf k}|^2}\ri) t^2.
\eea
Thus, even for the case of relativistic neutrinos,  QFT corrections
are in principle observable (for sufficiently  small time arguments).

 We also note  that the above
quantities are not interpreted as  probabilities, rather they
have a sense as {\em statistical averages}, i.e. as
mean values. This is because the structure of
the theory for mixed field is that of a many--body theory, where
does not make sense to talk about single--particle states. This
situation has a formal analogy with QFT
at finite temperature, where only statistical averages are well
defined.

We now show \cite{remarks} that the above results are consistent
with the generalization introduced in \S\ref{general}, i.e.  that the exact
oscillation probabilities are independent of the arbitrary mass
parameters.

It can be indeed explicitly checked that
\begin{equation}\label{miracle}
\langle {\widetilde \alpha}^r_{{\bf k},e}|
{\widetilde Q}_\sigma(t)  | {\widetilde\alpha}^r_{{\bf k},e}\rangle \, =
\, \langle  \alpha^r_{{\bf k},e}|
Q_\sigma(t) |\alpha^r_{{\bf k},e}\rangle
\end{equation}
which ensure the cancellation of the arbitrary mass parameters.

Note that the flavor charge operators $Q_\sigma(t)$ are {\em
invariant} under the action of the Bogoliubov generator
Eq.(\ref{FHYBVb}); however this is not sufficient to guarantee the
result Eq.(\ref{miracle}) which is non-trivial and provide a
criterion for the selection of the observables for mixed fields \cite{observables}.
As a matter of fact, the number operators for mixed fields are not
good observables since their expectation values do depend on the
arbitrary mass parameters. In \S\ref{neutral} we will consider
another observable, the momentum operator.

\subsection{Meson oscillations}
\label{mesosc}

The bosonic counterpart of the above oscillation formulas can be
derived in a similar way by use of the flavor charges for boson
fields \cite{bosonmix}.
By defining the mixed bosonic state as:
\be \label{bosonstate}
| a_{{\bf k},A}\ran \,\equiv\, a^{\dag}_{{\bf k},A}(0) \,
|0\ran_\AB
\ee
and the flavor charges ($\si= A, B$):
\bea
Q_\si(t) =  \int d^3 {\bf k}  \lf( a^{\dag}_{{\bf k},\si}(t)
a_{{\bf k},\si}(t)\,
-\, b^{\dag}_{-{\bf k},\si}(t)b_{-{\bf k},\si}(t)\ri)\,,
\eea
we obtain
${}_\AB\lan 0 | \,Q_\si(t)\, | 0\ran_\AB \,=\, 0 $
and
\bea\non
{\cal Q}_{{\bf k},\si}(t) &\equiv&
\lan a_{{\bf k},A} | \,Q_\si(t)\, | a_{{\bf k},A}\ran
\\[2mm]
&=&
\lf|\lf[a_{{\bf k},\si}(t), a^{\dag}_{{\bf k},A}(0) \ri]\ri|^2 \;
- \; \lf|\lf[b^\dag_{-{\bf k},\si}(t), a^{\dag}_{{\bf k},A}(0) \ri]\ri|^2.
\eea
The conservation of the total charge gives
$\sum_\si {\cal Q}_{{\bf k},\si}(t)\, = \,1$ and the oscillation formulas are:
\bea\non
{}\hspace{-.7cm}{\cal Q}_{{\bf k},A}(t)&=&
1\, - \,  \sin^2( 2 \te)   \,|U_{{\bf k}}|^2 \,
\sin^2 \lf( \frac{\om_{k,2} - \om_{k,1}}{2} t \ri)
\\[2mm] \label{bososc1}
  && {}\hspace{.4cm} +   \sin^2( 2 \te)   \,|V_{{\bf k}}|^2   \,
\sin^2 \lf( \frac{\om_{k,2} + \om_{k,1}}{2} t \ri)\,,
\\[3mm] \non
{\cal Q}_{{\bf k},B}(t) &=&
 \sin^2( 2 \te)
\, |U_{{\bf k}}|^2 \, \sin^2 \lf( \frac{\om_{k,2} - \om_{k,1}}{2} t \ri)
\\[2mm]  \label{bososc2}
&& -  \sin^2( 2 \te)
\, |V_{{\bf k}}|^2 \,
\sin^2 \lf( \frac{\om_{k,2} + \om_{k,1}}{2} t \ri)\,.
\eea
Thus also for bosons, the non-trivial structure of the flavor vacuum
induces corrections to the usual QM expressions for flavor oscillations.
The most obvious difference with respect to fermionic case is in the
negative sign which makes it possible a negative value for the bosonic
flavor charges. This only reinforces the statistical interpretation
given above, i.e. we are not dealing anymore with probabilities for
single particle evolution. As already noted for neutrinos,
in the relativistic limit the usual QM formulas are (approximately)
recovered.

\subsection[Mixing  and oscillations  of neutral particles]
{Mixing  and oscillations of neutral particles}
\label{neutral}

The above scheme is only valid for charged fields, since in the case of
neutral fermions (Majorana) and bosons, the (flavor) charges vanish
identically.
It is however possible to identify also in this case the relevant
observables for the description of flavor oscillations.

As an example, let us consider the case of a neutral boson field,
analogous treatment can be done for the Majorana field \cite{neutral}:
the notation is the same as in \S\ref{bosonmix}, the  mixing relations
being given by Eq.(\ref{bosmix}). The expansion for the
neutral field is (with $x_0\equiv t$):
\be\label{phimass}
\phi_{i}(x) \,=\,\int
\frac{d^3{\bf k}}{(2\pi)^{\frac{3}{2}}} \frac{1}{\sqrt{2\om_{k,i}}}
\lf( a_{{\bf k},i}\ e^{-i \om_{k,i} t } + a^{\dag }_{-{\bf k},i}\
e^{i \om_{k,i} t} \ri)e^{i {\bf k \cdot x}},
\ee
with $i=1,2$ and a similar expansion for the conjugate momenta
$\pi_{i}(x)$.
The generator of the mixing transformations can be
written as by $G_{\te}(t) = \exp[\te(S_{+}(t) -
S_{-}(t))]$ with
\bea \label{su2charges1}
&& {}\hspace{-1.5cm}
S_{+}(t) \equiv -i\;\int d^{3}{\bf x} \; \pi_{1}(x)\phi_{2}(x),
\quad
S_{-} (t)\equiv -i\;\int d^{3}{\bf x} \;\pi_{2}(x)\phi_{1}(x)
\\ [2mm] \label{su2charges2}
&& S_{3} \equiv \frac{-i}{2} \int d^{3}{\bf x}
\lf(\pi_{1}(x)\phi_{1}(x) - \pi_{2}(x)\phi_{2}(x)\ri)
\eea
The $SU(2)$ structure is thus still present, although being not related to
any flavor charges.

The flavor annihilation operators take now the following form \cite{neutral}:
\bea
a_{{\bf k},A}(t) \,=\,
\cos\te\;a_{{\bf k},1}\;+\;\sin\te\;
\lf( U_{{\bf k}}^*(t)\;a_{{\bf k},2}\;+
\; V_{{\bf k}}(t)\; a^{\dag}_{-{\bf k},2}\ri),
\\
a_{{\bf k},B}(t) \,=\,
\cos\te\;a_{{\bf k},2}\;-\;\sin\te\;
\lf( U_{{\bf k}}(t)\;a_{{\bf k},1}\;-
\; V_{{\bf k}}(t)\; a^{\dag}_{-{\bf k},1}\ri).
\eea
where the Bogoliubov coefficients coincide with those above
defined for charged bosons.

We then consider the momentum operator, defined as \cite{Itz}:
$P^j\equiv\int d^{3}{\bf x}\Te^{0j}(x)$, with
$\Te^{\mu\nu}\equiv\pa^\mu\phi\pa^\nu\phi -
g^{\mu\nu}\lf[\frac{1}{2}(\pa\phi)^2-\frac{1}{2}m^2\phi^2\ri]$.
For the free fields $\phi_i$ we have:
\bea\label{pmass}
&&{}\hspace{-1.3cm}
 {\bf P}_i\,=\,\int d^{3}{\bf x} \;\pi_{i}(x){\bf \nabla}\phi_{i}(x)
\, =\, \int d^{3}{\bf k} \, \frac{\bf k}{2}\,
\left(a_{{\bf k},i}^{\dag}a_{{\bf k},i} \, -\,
 a_{-{\bf k},i}^{\dag}a_{-{\bf k},i} \ri)\,,
\eea
with $ i=1,2$.
The momentum operator for mixed fields is:
\be\label{pflav}
{\bf P}_\si(t)\equiv G^{-1}_\te(t)\,{\bf P}_i\, G_\te(t)
 = \int d^{3}{\bf k} \, \frac{\bf k}{2}
\left(a_{{\bf k},\si}^{\dag}(t)a_{{\bf k},\si} (t) -
 a_{-{\bf k},\si}^{\dag}(t)a_{-{\bf k},\si}(t) \ri) ,
\ee
with $\si=A,B$.
Note that the total momentum
is conserved in time:
$ {\bf P}_A(t) \, + \, {\bf P}_B(t) \, =\,{\bf P}_1 \, +
\, {\bf P}_2 \, \equiv\, {\bf P}$.
Let us now consider the expectation values of the momentum operator for flavor fields
on the flavor state $|a_{{\bf k},A}\ran_\AB$, defined as in
Eq.(\ref{bosonstate}).
Obviously, this  is an eigenstate of ${\bf P}_A(t)$
at time $t=0$:
\bea
{\bf P}_A(0) \,|a_{{\bf k},A}\ran \, =\, {\bf k}\,|a_{{\bf k},A}\ran,
\eea
which follows from ${\bf P}_1 \,
|a_{{\bf k},1}\ran \, =\, {\bf k}\,|a_{{\bf k},1}\ran$ by
application of $G^{-1}_\te(0)$.

At time $t\neq 0$, the expectation value of the
momentum (normalized to the initial value) gives
${}_\AB\lan 0 | {\bf P}_\si(t) | 0\ran_\AB \, =\, 0$ and:
\bea\non {}\hspace{-1.3cm}{\cal P}_\si^{\bf k}(t)&\equiv&
\frac{\lan a_{{\bf k},A} | {\bf P}_\si(t) | a_{{\bf k},A}\ran}
{\lan a_{{\bf k},A} | {\bf P}_\si(0) |  a_{{\bf k},A}\ran} \,
\\[2mm]
&=& \lf|\lf[ a_{{\bf k},\si}(t),  a^{\dag}_{{\bf k},A}(0) \ri]\ri|^2 \;
- \; \lf|\lf[a^\dag_{-{\bf k},\si}(t), a^{\dag}_{{\bf k},A}(0) \ri]\ri|^2 \,,
\eea
with $\si=A,B$,
which is of the same form as the expression one obtains for the charged field.
The oscillation formulas coincide with those in
Eqs.(\ref{bososc1}),(\ref{bososc2}). Similar results are valid for Majorana
neutrinos \cite{neutral}.

\section{Geometric phase for oscillating particles}
\label{berry}

Let us now see how the notion of geometric phase
\cite{Anandan} enters the physics of mixing by considering the
example of neutrino oscillations.

We consider here
  two flavor mixing in the Pontecorvo approximation \cite{berry}, for
an extension to three flavors see Ref.\cite{berry3}.
The flavor states are:
\begin{eqnarray} \label{nue0a}
|\nu_{e}\rangle &=& \cos\theta\;|\nu_{1}\rangle \;+\;
\sin\theta\; |\nu_{2}\rangle \;
\\ [2mm] \label{nue0b}
|\nu_{\mu}\rangle &=& -\sin\theta\;|\nu_{1}\rangle \;+\;
\cos\theta\; |\nu_{2}\rangle \; .
\end{eqnarray}
The electron neutrino state at time $t$  is  \cite{Pont}
\begin{equation}\label{nue1}
{}\hspace{-.5cm}
|\nu_{e}(t)\rangle \equiv e^{-i H t} |\nu_{e}(0)\rangle= e^{-i
\omega_{1} t} \lf(\cos\theta\;|\nu_{1}\rangle \;+\; e^{-i
(\omega_{2}-\omega_{1}) t}\; \sin\theta\; |\nu_{2}\rangle \;
\ri),
\end{equation}
where $H |\nu_i\rangle = \omega_i |\nu_i\rangle$, $i=1,2$.
The state $|\nu_{e}(t)\rangle$, apart from a phase factor,
reproduces the initial state $|\nu_{e}(0)\rangle$ after a period
$T= \frac{2\pi}{\omega_{2} - \omega_{1}}$ :
\begin{equation}\label{nue2}
|\nu_{e}(T)\rangle = e^{i \phi} |\nu_{e}(0)\rangle
\;\;\;\;\;\;\;\;\;\;\;\;,\;\;\;\;\;\;\;\;\;\;\;\;\; \phi= -
\frac{2\pi \omega_{1}}{\omega_{2} - \omega_{1}} \,.
\end{equation}
We now show how such a time evolution does contain a purely
geometric part. It is  straightforward
 to separate the geometric and dynamical phases
following the standard procedure \cite{Anandan}:
\begin{eqnarray}\nonumber
&&{}\hspace{-1cm}
\beta_e\,=\, \phi + \int_{0}^{T} \;\langle \nu_{e}(t)|\;
i\partial_t\;|\nu_{e}(t)\rangle \,dt
\\ \label{ber1}
&&{}\hspace{-1cm}
\,=\,- \frac{2\pi \omega_{1}}{\omega_{2} - \omega_{1}} +
\frac{2\pi}{\omega_{2} - \omega_{1}}(\omega_{1}\;\cos^2\theta +
\omega_{2}\;\sin^2\theta)\;= \; 2 \pi \sin^{2}\theta  \, .
\end{eqnarray}
We thus see that there is indeed a non-zero geometrical phase
$\beta_e$,
 related to the mixing angle
$\theta$, and that it is independent from the neutrino energies
 $\omega_i$ and masses $m_i$.
In a similar fashion, we obtain the Berry phase for the muon
neutrino state:
\begin{equation}\label{ber1b}
\beta_{\mu}\,= \,\phi + \int_{0}^{T} \;\langle \nu_{\mu}(t)|\;
i\partial_t\;|\nu_{\mu}(t)\rangle \,dt \, =\, 2 \pi \cos^{2}\theta
\, .
\end{equation}
Note that $\beta_e + \beta_{\mu} = 2\pi$.

Generalization to $n-$cycles is also interesting. Eq.(\ref{ber1})
can be rewritten for the  $n-$cycle case as
\begin{equation}\label{ber1n}
\beta^{(n)}_{e}\,= \, \int_{0}^{nT} \;\langle \nu_{e}(t)|\;
\lf(i\partial_t -\omega_1\ri)\;|\nu_{e}(t)\rangle \,dt \, =\, 2 \pi
\,n\,\sin^{2}\theta \, ,
\end{equation}
Eq.(\ref{ber1n}) shows that the Berry phase acts
as a ``counter'' of neutrino oscillations, adding up $2 \pi
\,\sin^{2}\theta$ to the phase of the (electron) neutrino state
after each complete oscillation.

In Ref.\cite{berry}, a gauge
structure  and a covariant derivative were introduced
in connection with the above geometric structures.

The case of three
flavor mixing has been analyzed in Ref.\cite{berry3}.
The above result also applies to other (similar) cases
of particle oscillations, for example to Kaon oscillations.
Finally, we
note  that a measurement of the above geometric phase would give
a direct measurement of the mixing angle independently from the
values of the masses.


\section{Three flavor fermion mixing}
\label{3flav}

We now consider some aspects of fermion mixing in the case of three
flavors \cite{BV95,3flavors}.
This is particularly relevant
 because of the possibility of $CP$ violation associated with
it.
Among the various possible parameterizations of the mixing matrix
for three fields, we choose to work with the standard representation of the CKM matrix
\cite{Cheng-Li}:
\bea\label{fermix}
&&{}\hspace{-1cm}
\Psi_f(x) \, = {\cal U} \, \Psi_m (x)
\\
&&{}\hspace{-1cm} \non
{\cal U}=\lf(\begin{array}{ccc}
c_{12}c_{13} & s_{12}c_{13} & s_{13}e^{-i\de} \\
-s_{12}c_{23}-c_{12}s_{23}s_{13}e^{i\de} &
c_{12}c_{23}-s_{12}s_{23}s_{13}e^{i\de} & s_{23}c_{13} \\
s_{12}s_{23}-c_{12}c_{23}s_{13}e^{i\de} &
-c_{12}s_{23}-s_{12}c_{23}s_{13}e^{i\de} & c_{23}c_{13}
\end{array}\ri),
\eea
with $c_{ij}=\cos\te_{ij}$ and  $s_{ij}=\sin\te_{ij}$, being
$\te_{ij}$ the mixing angle between $\nu_{i},\nu_{j}$ and
$\Psi_m^T=(\nu_1,\nu_2,\nu_3)$,
$\Psi_f^T=(\nu_e,\nu_\mu,\nu_\tau)$.

As shown in  Ref.\cite{BV95}, the generator of the
 transformation (\ref{fermix}) is:
\bea\label{incond}
&&\nu_{\si}^{\al}(x)\equiv G^{-1}_{\bf \te}(t)
\, \nu_{i}^{\al}(x)\, G_{\bf \te}(t), \eea
with $(\si,i)=(e,1), (\mu,2), (\tau,3)$, and
\bea\label{generator} &&G_{\bf
\te}(t)=G_{23}(t)G_{13}(t)G_{12}(t)\, , \eea
where $ G_{ij}(t)\equiv \exp\Big[\te_{ij}L_{ij}(t)\Big]$ and
\bea\label{generators1}
&&\hspace{-1cm} L_{12}(t)=\int
d^3{\bf x}\lf[\nu_{1}^{\dag}(x)\nu_{2}(x)-\nu_{2}^{\dag}(x)\nu_{1}(x)\ri],
\\ [2mm] \label{generators2}
&&\hspace{-1cm}L_{23}(t)=\int
d^3{\bf x}\lf[\nu_{2}^{\dag}(x)\nu_{3}(x)-\nu_{3}^{\dag}(x)\nu_{2}(x)\ri],
\\[2mm] \label{generators3}
&&\hspace{-1.4cm}L_{13}(\de,t)=\int
d^3{\bf x}\lf[\nu_{1}^{\dag}(x)\nu_{3}(x)e^{-i\de}-\nu_{3}^{\dag}(x)
\nu_{1}(x)e^{i\de}\ri]. \eea
It is evident from the above form of the generators, that the
phase $\de$ is unavoidable for three field mixing, while it can be
incorporated in the definition of the fields in the two flavor
case.

In Ref.\cite{3flavors}, the flavor vacuum and the flavor annihilation operators were
studied for the above mixing relations. Oscillation formulas were derived exhibiting
$CP$ violation.
Here we do not report on these results, rather we comment on the algebraic structure
associated with the generator Eq.(\ref{generator}).
Indeed, the generators Eqs.(\ref{generators1})-(\ref{generators3}) can be obtained
by acting on the triplet $\Psi_m^T=(\nu_1,\nu_2,\nu_3)$ with global phase
transformations, in analogy with what has been done in  \S\ref{fermcurr}. One then
obtains the following set of charges \cite{3flavors}:
\bea\label{su3charges}
&&{}\hspace{-.5cm}
{\ti Q}_{m,j}(t)\, =\, \int  d^3{\bf x}\,\Psi^\dag_m(x)\, {\ti F}_j\,
\Psi_m(x) \;, \qquad j=1, 2,..., 8.
\eea
where ${\ti F}_{j}\equiv\frac{1}{2}{\ti \lambda}_{j}$ and
the ${\ti \lambda}_{j}$ are a generalization of the usual
Gell-Mann matrices $\lambda_{j}$:
\bea \non
&&{}\hspace{-.5cm}
{\ti \lambda}_{1}=\lf(\ba{ccc}
  0 & e^{i\de_2} & 0 \\
 e^{-i\de_2} & 0 & 0 \\
  0 & 0 & 0
\ea\ri)
\;,\qquad
{\ti \lambda}_{2}=\lf(\ba{ccc}
  0 & -i e^{i\de_2} & 0 \\
  i e^{-i\de_2}& 0 & 0 \\
  0 & 0 & 0
\ea \ri)
\\[2mm] \non
&&{}\hspace{-.5cm}
{\ti \lambda}_{4}=\lf(\ba{ccc}
  0 & 0 & e^{-i\de_5}\\
  0 & 0 & 0 \\
  e^{i\de_5} & 0 & 0
\ea\ri)\;,\qquad
{\ti \lambda}_{5}=\lf(\ba{ccc}
  0 & 0 & -ie^{-i\de_5} \\
  0 & 0 & 0 \\
  ie^{i\de_5} & 0 & 0
\ea\ri)
\\ [2mm] \non
&&{}\hspace{-.5cm}
{\ti \lambda}_{6}=\lf(\ba{ccc}
  0 & 0 & 0 \\
  0 & 0 & e^{i\de_7} \\
  0 & e^{-i\de_7} & 0
\ea\ri)\;, \qquad
{\ti \lambda}_{7}=\lf(\ba{ccc}
  0 & 0 & 0 \\
  0 & 0 & -i e^{i\de_7}\\
  0 & ie^{-i\de_7} & 0
\ea\ri)
 \;,
\\[2mm] \label{gelm}
&&{}\hspace{-.5cm}
{\ti \lambda}_{3}=\lf(\ba{ccc}
  1 & 0 & 0 \\
  0 & -1 & 0 \\
  0 & 0 & 0
\ea\ri)\;, \qquad
{\ti \lambda_{8}}=\frac{1}{\sqrt{3}}\lf(\ba{ccc}
  1 & 0 & 0 \\
  0 & 1 & 0 \\
  0 & 0 & -2
\ea\ri).
\eea
These are normalized as  $tr(  {\ti \lambda}_{j}
{\ti \lambda}_{k}) =2\delta_{jk}$.
Thus the matrix Eq.(\ref{fermix}) is generated by ${\ti Q}_{m,2}(t)$,
${\ti Q}_{m,5}(t)$ and ${\ti Q}_{m,7}(t)$, with $\{\de_2,\de_5,\de_7\}\rar
\{0,\de,0\}$.

The interesting point is that the algebra generated by the matrices
Eq.(\ref{gelm}) {\em is not} $su(3)$ unless the condition
$\De \equiv \de_2+\de_5 +\de_7 =0$ is imposed:
such a condition is however incompatible
with the presence of a CP violating phase.
When CP violation is allowed, then $\De \neq 0$ and the $su(3)$ algebra
is deformed.
To see this, let us introduce the
raising and lowering operators, defined as \cite{Cheng-Li}:
\bea &&{}\hspace{-1cm}
{\ti T}_\pm \equiv {\ti F}_1 \pm i {\ti F}_2 \quad, \quad
{\ti U}_\pm \equiv {\ti F}_6 \pm i {\ti F}_7 \quad, \quad {\ti
V}_\pm \equiv {\ti F}_4 \pm i {\ti F}_5 \eea
We also define:
\bea &&{}\hspace{-1.4cm}
{\ti T}_3 \equiv  {\ti F}_3 \quad, \quad {\ti U}_3 \equiv
\frac{1}{2}\lf(\sqrt{3} {\ti F}_8 - {\ti F}_3 \ri) \quad, \quad
{\ti V}_3 \equiv \frac{1}{2}\lf(\sqrt{3} {\ti F}_8 + {\ti F}_3
\ri) \eea
Then the deformed commutators are the following ones:
\bea \non
&&{}\hspace{-1.5cm}
[{\ti T}_+,{\ti V}_-] \,=\, - {\ti U}_- \,e^{2 i \De {\ti U}_3}
\quad, \quad [{\ti T}_+,{\ti U}_+] \,=\,  {\ti V}_+ \,e^{-2 i \De
{\ti V}_3} \;,
\\ [2mm] && {}\hspace{.5cm}[{\ti U}_+,{\ti V}_-] \,=\,  {\ti T}_-
\,e^{2 i \De {\ti T}_3}\,,
\eea
all the others being identical to the ordinary $su(3)$
ones \cite{Cheng-Li}.

\section{Neutrino oscillations from relativistic flavor current}
\label{spacemix}

A realistic description of neutrino oscillations requires to take
into account the finite size of source and detector and the fact that
 in current experiments what is measured is the distance source-detector
rather than the time of flight of (oscillating) neutrinos.
Thus various approaches were developed, based on wave-packets and leading
to a space-dependent oscillation formula \cite{Kayser}-\cite{Ishikawa}.

Here we report about recent results, showing how an exact expression for
QFT space-dependent oscillation formula can be found by using the above
defined flavor states and relativistic flavor currents \cite{BPT02}. Such
an approach was first proposed in Ref.\cite{Anco} in the context of
non-relativistic QM (see also Ref.\cite{Zralek}).
We thus consider the flux of (electron) neutrinos through a detector surface
\bea
\Phi_{\nu_e\to \nu_e}(L)=\int_{t_0}^{T}
d t\int_{\Omega} \langle \nu_e| J_e^{i}(\bx,t)|
\nu_e\rangle\,\, d {\bf S}^i
\eea
The  neutrino state is described by a wave packet:
\bea
| \nu_e(\bx_0,t_0)\rangle=A\int d^3
\bk \,  {e}^{-i(\om_{k,1}t_0-\bk\cdot\bx_0)}
f(\bk)\,\alpha_{\bk,e}^{r\dag}(t_0) \,|0(t_0)\rangle_{e,\mu}
\eea
The flavor current is:
$J^\mu_e (x) = \bar{\nu}_e(x) \gamma^{\mu} \nu_e(x)$.
In Ref.\cite{BPT02} it is shown that
$_{e,\mu}\langle0|J^{\mu}(\bx,t)|0\rangle_{e,\mu}=0$ and
\bea\label{J1}
&&\langle \nu_e |J_e^\mu(\bx,t)| \nu_e\rangle
\,=\, {\bar \Psi}(\bx,t)\, \Ga^\mu \lf(\begin{array}{cc}1&1\\1&1\end{array}\ri)
\,\Psi(\bx,t)
\eea
with
\bea
&&{}\hspace{-.5cm}
\Psi(\bx,t) \equiv A \int\frac{d^3 \bk}{(2\pi)^\frac{3}{2}}\,
e^{i\bk \cdot \bx}\,f(\bk)\,
\lf(\begin{array}{l} u_{\bk,1}^r \, X_{\bk,e}(t)
\\ [3mm]
 \sum_{s} v_{-\bk,1}^s \,(\vec{\si}\cdot \bk)^{sr}\, Y_{\bk,e}(t)
 \end{array} \ri)
\\ [2mm] \non
&&{}\hspace{-.5cm}
X_{\bk,e}(t)\,=\,\cos^2\theta e^{-i\om_{k,1}t}+\sin^2\theta
\left[e^{-i\om_{k,2} t}|U_{\bk}|^2+
e^{i \om_{k,2}t}|V_{\bk}|^2\right]
\,
\\[2mm] \non
&&{}\hspace{-.5cm}
Y_{\bk,e}(t) \,=\,\sin^2\theta|U_{\bk}|\chi_1\chi_2
\left[\frac{1}{\om_{k,2}+m_2}  -
\frac{1}{\om_{k,1}+m_1}\right]\left[e^{-i \om_{k,2} t}-
e^{i \om_{k,2}t}\right]
\eea
where  $\vec{\sigma} \cdot \bk  =
\lf(\begin{array}{cc} k_3 & k_-\\ k_+ & -k_3\end{array}\ri)$
and $\chi\,_{i}\equiv
\left(\frac{\omega_{k,i}+m_{i}}{4 \omega_{k,i}}\right)^{\frac{1}{2}}$.

The expression in Eq.(\ref{J1}) contains the most general
information about neutrino oscillations and can be explicitly
evaluated  once the form of the wave-packet is specified. A
similar expression can be easily obtained for the other quantity
of interest, namely $\langle \nu_e |J_\mu^{\,{\mu}}(\bx,t)|
\nu_e\rangle$.

An oscillation formula in space is then obtained in Ref.\cite{BPT02} in
the case of spherical symmetry and by assuming a
gaussian wave packet for the flavor state:
\bea
f_k=\frac{1}{(2\pi\si_k^2)^{\frac{1}{4}}}\,\,
\exp\left[-\frac{(k-Q)^2}{4\sigma^2_{k}}\right]
\eea
Such an expression can be evaluated numerically (see Fig.(\ref{fig3})) and
it reduces \cite{BPT02} to the standard
formula \cite{giuntipak,Beuthe} in the relativistic limit:
\bea
&&{}\hspace{-.7cm}
\Phi_{\nu_e\to \nu_e}(z)\simeq  1-\frac{1}{2}\sin^2(2\theta)
\\ \non
&& {}\hspace{.5cm}+\frac{1}{2}\sin^2(2\theta)
\cos\left(2\pi\frac{z}{L^{osc}}\right)
\,\exp\lf[-\left(\frac{z}{L^{coh}}\right)^2  -2\pi^2
\left(\frac{\sigma_x}{L^{osc}}\right)^2\ri]
\eea
with $L_{osc}=\frac{4\pi Q}{\De m^2}$ and
$ L_{coh}=\frac{L_{osc}Q}{\sqrt{2}\pi\sigma_{k}}$ being the
usual oscillation
length and  coherence length \cite{giuntipak,Beuthe}.

\begin{figure}[t]
\setlength{\unitlength}{1mm} \vspace*{70mm} 
\includegraphics{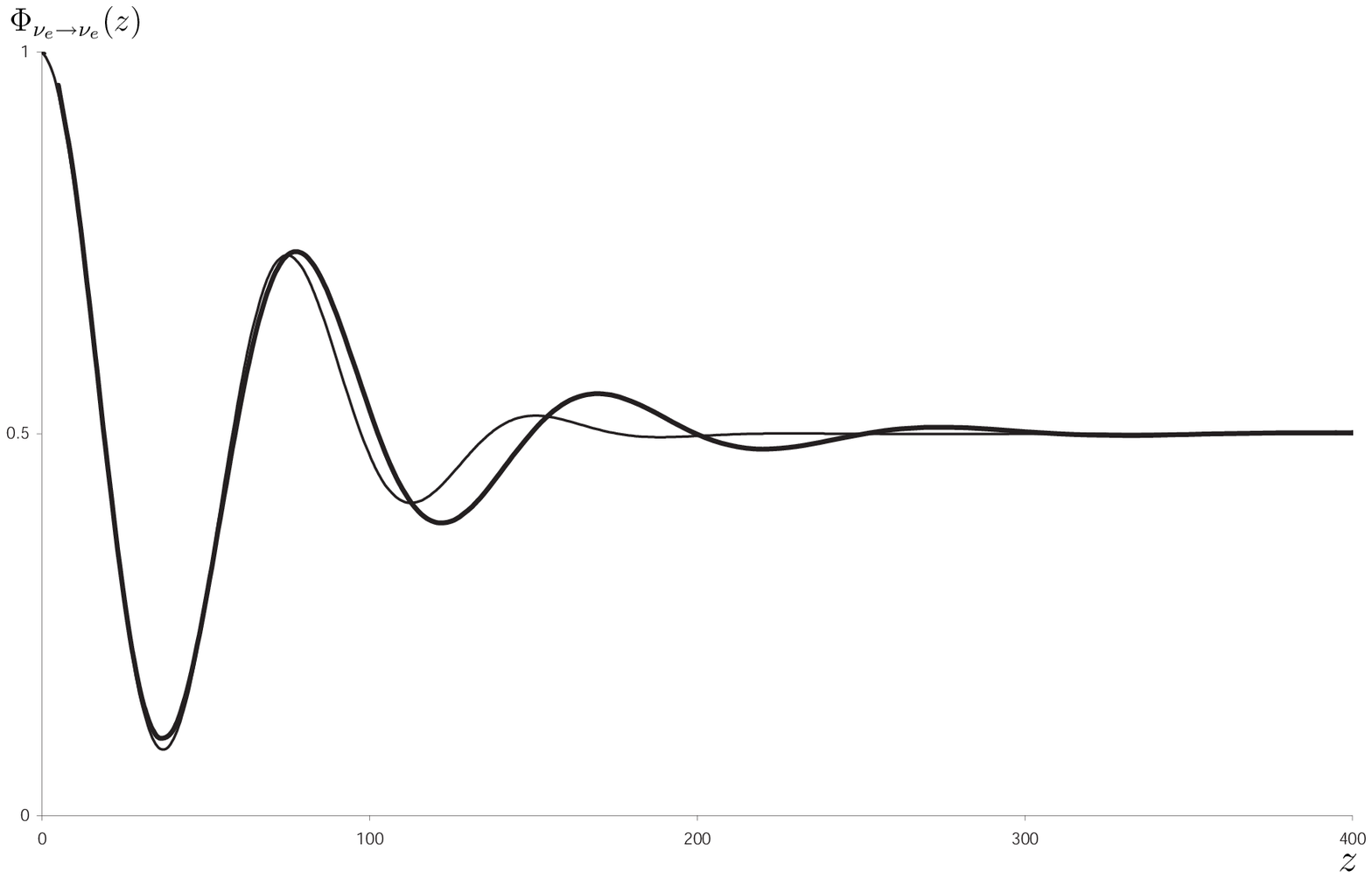}
\caption{QFT flux (thick line) vs. standard
 formula (thin line)
for $\te=\pi/4$, $\si_k =10$, $m_\1=1$, $m_\2=3$, $Q=50$.}

\mlab{fig3}
\end{figure}

\section{Summary}
In this report we have discussed recent results in the area of
field mixing and oscillations. We have shown that a consistent
field theoretical treatment is possible, both for
fermions and for bosons, once we realize the
unitary inequivalence of the mass and flavor representations.
The flavor Hilbert space is thus constructed and the flavor
vacuum is shown to have the structure  of a $SU(n)$ generalized
coherent state, for the case of mixing among $n$ generations\footnote{When no
$CP$ violating phases are present - see \S\ref{3flav}.}.
We have then discussed the algebraic structure of the currents
and charges associated with field mixing.

On the basis of these results, exact oscillation formulas have
been calculated, exhibiting non-perturbative corrections with
respect to the usual QM ones. The usual formulas are shown
to be approximately valid in
the relativistic region. Exact oscillation formulas in space can
also be derived by use of the relativistic flavor currents.

We have also shown that a geometric phase is associated to
flavor oscillations and discussed the role of the CP violating
phase in connection with the algebra of currents associated
to three flavor mixing.

For lack of space, we have omitted other interesting development,
in particular we would like to mention the analysis, in the above
framework, of the neutrino oscillations in matter (MSW effect) \cite{msw}.
An interesting new line of research is the investigation of the
issue of Lorentz invariance for the flavor states \cite{dispersion}:
deformed dispersion relations for neutrino
flavor states may
be indeed incorporated into frameworks encoding the breakdown of Lorentz
invariance \cite{leejoao}.

\acknowledgements
We acknowledge  the ESF Program COSLAB, EPSRC,
INFN and INFM for partial financial support.

\end{article}

\begin{thebibliography}{99}


\bibitem{Cheng-Li}
T.~Cheng and L.~Li, {\it Gauge Theory of Elementary Particle
Physics}, Clarendon Press, Oxford, 1989.



%
\bibitem{experiments}
J.~Davis, D.~S.~Harmer and K.~C.~Hoffmann,
Phys.\ Rev.\ Lett. {\bf 20} (1968) 1205.
%
M.~Koshiba,
in  ``Erice 1998, From the Planck length to the
Hubble radius'', 170; S.~Fukuda et al. [Super-Kamiokande
collaboration],
Phys.\ Rev.\ Lett. {\bf 86} (2001) 5656.
%
Q.~R.~Ahmad et al. [SNO collaboration]
Phys.\ Rev.\ Lett. {\bf 87} (2001) 071301;
Phys.\ Rev.\ Lett.\  {\bf 89} (2002) 011301.
%
K.~Eguchi {\it et al.}  [KamLAND Collaboration],
Phys.\ Rev.\ Lett.\  {\bf 90} (2003) 021802
%
M.~H.~Ahn {\it et al.}  [K2K Collaboration],
Phys.\ Rev.\ Lett.\  {\bf 90} (2003) 041801


\bibitem{Pont}   B.~Pontecorvo,  Zh.\ Eksp.\ Theor.\ Fiz. {\bf 33} (1958)
549; JEPT {\bf 6} (1958) 429; Z.~Maki, M.~Nakagawa and
S.~Sakata, Prog.\ Theor.\ Phys. {\bf 28} (1962) 870;
V.~Gribov and
B.~Pontecorvo, Phys.\ Lett. \ B {\bf 28} (1969)  493;
S.~M.~Bilenky and
B.~Pontecorvo, Phys.\ Rep. {\bf 41} (1978) 225.


%
\bibitem{Sol}
R.~Mohapatra and P.~Pal, {\it Massive Neutrinos in Physics and
Astrophysics}, (World Scientific, Singapore, 1991);
J.~N.~Bahcall, {\it Neutrino Astrophysics}, (Cambridge Univ. Press,
Cambridge, 1989).

\bibitem{Barg}
V.~Bargmann,
Annals Math.\  {\bf 59} (1954) 1;
A.~Galindo and P.~Pascual, {\it Quantum Mechanics}, (Springer
Verlag, 1990).

\bibitem{Greenb}
D.~M.~Greenberger, Phys. \ Rev.\ Lett. {\bf 87}  (2001) 100405.


\bibitem{Zralek}
M.~Zralek,
Acta Phys.\ Polon.\  B {\bf 29} (1998) 3925.


\bibitem{kimbook}
C.~Giunti, C.~W.~Kim and U.~W.~Lee,
Phys.\ Rev.\ D {\bf 45} (1992) 2414;
C.W. Kim and A. Pevsner, {\it Neutrinos in Physics and
Astrophysics}, Harwood
Academic Press, Chur, Switzerland,1993.

%
%
\bibitem{BV95}
M.~Blasone and G.~Vitiello,
Annals Phys.\  {\bf 244} (1995) 283 [Erratum-ibid.\  {\bf 249}
(1995) 363].

\bibitem{lathuile}
M.~Blasone, P.~A.~Henning and G.~Vitiello, in M.Greco Ed.``La Thuile 1996,
Results and perspectives in particle physics'', INFN
Frascati 1996, p.139-152 [hep-ph/9605335].

\bibitem{BHV99}
M.~Blasone, P.~A.~Henning and G.~Vitiello,
Phys.\ Lett.\ B {\bf 451} (1999) 140;
M.~Blasone, in A.Zichichi Ed. ``Erice 1998, From the Planck length to the Hubble
radius'' (World Scientific) p.584,
[hep-ph/9810329].

\bibitem{binger}
M.~Binger and C.~R.~Ji,
Phys.\ Rev.\ D {\bf 60} (1999) 056005.
%
C.~R.~Ji and Y.~Mishchenko,
Phys.\ Rev.\ D {\bf 64} (2001) 076004;
%
Phys.\ Rev.\ D {\bf 65} (2002) 096015.

\bibitem{fujii1}
K.~Fujii, C.~Habe and T.~Yabuki,
Phys.\ Rev.\ D {\bf 59} (1999) 113003 [Erratum-ibid.\ D {\bf 60}
(1999) 099903];
Phys.\ Rev.\ D {\bf 64} (2001) 013011.

\bibitem{hannabuss}
K.~C.~Hannabuss and D.~C.~Latimer,
J.\ Phys.\ A  {\bf 36} (2003) L69;
J.\ Phys.\ A {\bf 33} (2000) 1369.

\bibitem{remarks}
M.~Blasone and G.~Vitiello,
Phys.\ Rev.\ D {\bf 60} (1999) 111302.

\bibitem{currents}
M.~Blasone, P.~Jizba and G.~Vitiello,
Phys.\ Lett.\ B {\bf  517 } (2001) 471.


\bibitem{comment}
M.~Blasone, A.~Capolupo and G.~Vitiello,
in Yue-Liang Wu, editor, {\it Flavor Physics}, 425-433. World Scientific,
Singapore 2002. [hep-th/0107125];

\bibitem{bosonmix}
M.~Blasone, A.~Capolupo, O.~Romei and G.~Vitiello,
Phys.\ Rev.\ D {\bf 63} (2001) 125015.

\bibitem{3flavors}
M.~Blasone, A.~Capolupo and G.~Vitiello,
Phys.\ Rev.\ D {\bf 66} (2002) 025033;


\bibitem{neutral}
M.~Blasone and J.~S.~Palmer, [hep-ph/0305257]

\bibitem{BPT02}
M.~Blasone, P.~P.~Pacheco and H.~W.~Tseung,
Phys.\ Rev.\ D {\bf 67} (2003) 073011.

\bibitem{observables}
M.~Blasone, P.~Jizba and G.~Vitiello,
[hep-ph/0308009].


\bib{Itz}
C.Itzykson and J.B.Zuber, {\it Quantum Field Theory},
(McGraw-Hill, New York, 1980);

%
\bib{Um1}
H.Umezawa,{\em Advanced Field Theory: Micro, Macro and Thermal
Physics} (American Institute of Physics, 1993)

\bibitem{Per}
A.~Perelomov, {\it Generalized Coherent States and Their
Applications}, (Springer--Verlag, Berlin, 1986).


\bibitem{berry}
M.~Blasone, P.~A.~Henning and G.~Vitiello, Phys.\ Lett.\  B {\bf
466} (1999) 262;

\bibitem{Anandan}
Y.~Aharonov and J.~Anandan {\it Phys. Rev. Lett.} {\bf 58} (1987)
1593; {\it Phys. Rev. Lett.} {\bf 65} (1990)
1697.
%

\bibitem{berry3}
X.~B.~Wang, L.~C.~Kwek, Y.~Liu and C.~H.~Oh,
Phys.\ Rev.\ D {\bf 63} (2001) 053003.










\bibitem{Kayser}
B.~Kayser,
Phys.\ Rev. \ D {\bf 24}  (1981) 110;
B.~Kayser, F.~Gibrat-Debut and D.~Perrier, { \it The
Physics of massive neutrinos,} World Scientific, 1989.


\bibitem{Beuthe}
M.~Beuthe,
Phys.\ Rev.\ D {\bf 66} (2002) 013003;
Phys.\ Rep.\  {\bf 375} (2003) 105.

\bibitem{giuntipak}

C.~Giunti,
JHEP {\bf 0211} (2002) 017.
C.~Giunti and C.~W.~Kim,
Phys.\ Rev.\ D {\bf 58} (1998) 017301.

\bibitem{Grimus}
W.~Grimus and P.~Stockinger,
Phys.\ Rev.\ D {\bf 54} (1996) 3414;
W.~Grimus, P.~Stockinger and S.~Mohanty,
Phys.\ Rev.\ D {\bf 59} (1999) 013011.

\bibitem{Cardall}
C.~Y.~Cardall,
Phys.\ Rev.\ D {\bf 61} (2000) 073006;
C.~Y.~Cardall and D.~J.~H.~Chung,
Phys.\ Rev.\ D {\bf 60} (1999) 073012.

\bibitem{Kiers}
K.~Kiers and N.~Weiss,
Phys.\ Rev.\ D {\bf 57} (1998) 3091.

\bibitem{Ishikawa}
T.~Yabuki and K.~Ishikawa, Prog.\ Theor.\ Phys.\ {\bf 108} (2002), 347.

\bibitem{Anco}
B.~Ancochea, A.~Bramon, R.~Munoz-Tapia and M.~Nowakowski,
Phys.\ Lett.\ B {\bf 389} (1996) 149.

\bibitem{msw}
K.~Fujii, C.~Habe and M.~Blasone,
[hep-ph/0212076].

\bibitem{dispersion}
M.~Blasone, J.~Magueijo and P.~Pires-Pacheco,
[hep-ph/0307205].

\bibitem{leejoao}
J.~Magueijo and L.~Smolin, Phys. \ Rev.\ D {\bf 67} (2003) 044017;
Phys.\ Rev.\ Lett. {\bf 88} (2002) 190403.
\end{thebibliography}
\end{document}